\documentclass[aps,pre,twocolumn,superscriptaddress]{revtex4}
\usepackage[english]{babel}
\usepackage[dvips]{graphicx}
\usepackage[dvips,unicode,colorlinks,linkcolor=blue,citecolor=blue,urlcolor=blue]{hyperref}
\usepackage{amssymb}
\usepackage{dcolumn}
\usepackage{bm}
\usepackage{amsmath}

\begin{document}

\title{Nonstationary heat conduction in one-dimensional models with substrate potential}

\author{O. V. Gendelman}

\affiliation{
Faculty of Mechanical Engineering, Technion -- Israel Institute of Technology,
Haifa 32000, Israel}

\affiliation{
contacting author, ovgend@tx.technion.ac.il
}

\author{R. Shvartsman}

\affiliation{
Faculty of Mechanical Engineering, Technion -- Israel Institute of Technology,
Haifa 32000, Israel}

\author{B. Madar}

\affiliation{
Faculty of Mechanical Engineering, Technion -- Israel Institute of Technology,
Haifa 32000, Israel}

\author{A. V. Savin}

\affiliation{Semenov Institute of Chemical Physics, Russian Academy
of Sciences, Moscow 117977, Russia}

\begin{abstract}
The paper investigates non-stationary heat conduction in one-dimensional models with
substrate potential. In order to establish universal characteristic properties of the process,
we explore three different models  --- Frenkel-Kontorova (FK), phi4+ ($\phi^4$+) and phi4-
($\phi^4-$). Direct numeric simulations reveal in all these models a crossover from oscillatory
decay of short-wave perturbations of the temperature field to smooth diffusive decay of the
long-wave perturbations. Such behavior is inconsistent with parabolic Fourier equation of the
heat conduction and clearly demonstrates the necessity of hyperbolic models. The crossover
wavelength decreases with increase of average temperature. The decay patterns of the temperature
field almost do not depend on the amplitude of the perturbations, so the use of linear evolution
equations for temperature field is justified. In all model investigated, the relaxation of thermal
perturbations is exponential -- contrary to linear chain, where it follows a power law.
However, the most popular lowest-order hyperbolic generalization of the Fourier law, known as
Cattaneo-Vernotte (CV) or telegraph equation (TE) is not valid for description of the observed behavior
of the models with on-site potential. In part of the models a characteristic
relaxation times  exhibit peculiar scaling with respect to the temperature perturbation wavelength.
Quite surprisingly, such behavior is similar to that of well-known model with divergent heat
conduction (Fermi-Pasta-Ulam chain) and rather different from the model with normal heat
conduction (chain of rotators). Thus, the data on the non-stationary heat conduction in the
systems with on-site potential do not fit commonly accepted concept of universality classes for
heat conduction in one-dimensional models.

\end{abstract}

\pacs{44.10.+i, 05.45.-a, 05.60.-k, 05.70.Ln}

\maketitle

\section{Introduction}

Relationship between empiric equations of heat conduction (Fourier law) and microstructure of
solid dielectrics is known to be one of the oldest and most elusive unsolved problems in solid
state physics \cite{p1,p2}. The classic solution for the problem suggested by Peierls \cite{p3}
was questioned after seminal numeric experiment of Fermi, Pasta and Ulam \cite{p4}; the latter
has disproved common belief concerning fast thermalization and mixing in non-integrable systems
with weak nonlinearity. Despite large amount of work done, necessary and sufficient conditions
for microscopic model of solid to obey macroscopically the Fourier law with finite and size --
independent heat conduction coefficient \cite{p4,p5, p6,p7,p8,p9,p10,p11,p12,p13,p14,p15,p16}
are not known yet. Numerous anomalies in the heat transfer in microscopic models of dielectrics
were revealed by means of direct numeric simulation over recent years, including qualitatively
different behavior of models of different types (with and without on-site potential) and
dimensionality \cite{p1}. To date, it is believed that in one dimension the microscopic models
with conserved momentum the heat conduction coefficient diverges in the thermodynamic limit
(as the chain length N goes to infinity) as $\kappa\sim N^\gamma$ with $\gamma$ varying in
the interval $0.3\div 0.4$ \cite{p1}. The only known exception with convergent heat conduction
coefficient is the chain of coupled rotators \cite{p8, p9}. As for the models without the moment
conservation, their heat conduction is convergent \cite{p10, p14}, with exception of integrable
models \cite{p11}. In two dimensions, the divergence still exists \cite{p1}, although is reported
to be of logarithmic rather than power-like type. It is believed that in three dimensions the
heat conduction coefficient will converge, although some alternative data exist \cite{p15,p16}.

Vast majority of results obtained to date in the field of microscopic foundations of the heat
conduction dealt with stationary problem, with steady heat flux under constant thermal gradient.
In all these cases the macroscopic law to be checked was standard Fourier law of heat conduction.
However, it is well-known that due to its parabolic character it implies infinite speed of the
signal propagation and thus should be modified if very large gradients or extremely small scales
are involved \cite{p17,p18,p19,p20}.
On macroscopic level, numerous modifications were suggested to recover the hyperbolic character
of the heat transport equation \cite{p19,p20,p21,p22}. Perhaps, the most known is the lowest
-- order approximation known as Cattaneo-Vernotte (CV) or telegraphist (TE) law \cite{p23,p24}.
It is written as
\begin{equation}
\label{f1}
(1+\tau \frac{\partial }{\partial t} )\vec{q}=-\kappa \nabla T
\end{equation}
where $\kappa$ is standard heat conduction coefficient and $\tau$ is characteristic relaxation
time of the system, $\vec{q}$ is the vector of heat flux and $T$ is the temperature.
The relaxation time is rather short for majority of materials ($10^{-12}$ s at room temperature),
but can be of order 1 s, for instance in some biological tissues \cite{p25}.

Modifications of the Fourier law bring about new observable physical phenomena, such as
temperature waves or the second sound \cite{p17,p18,p19,p20}. Importance of the hyperbolic heat
conduction models for description of a nanoscale heat transfer
has been recognized in recent experiments \cite{p26,p27}.

Only few papers dealt with microscopic foundations of non-Fourier heat conduction laws from
the first principles or by its numeric investigation in the microscopic models
\cite{p26, p28,p29,p30,p31,p32,p33,p34,p35,p36,p37}. These works confirm that the non-Fourier
effects may be very significant if large space gradients or fast changes of the temperature
are involved.

In recent paper \cite{p38} an attempt was made to relate the structure and parameters of
microscopic models with conserved momentum to empiric description of the non-Fourier heat conduction.
For this sake, two models belonging to different universality classes with respect to the
stationary heat conduction were studied: the $\beta$-FPU chain and the chain of rotators.
Oscillatory behavior of the decaying temperature disturbance, compatible only with hyperbolic
macroscopic equation of the thermal transport, has been revealed in both models. The CV equation
of the heat transport adequately describes the behavior of the chain of rotators, besides the
region of crossover between the "hyperbolic" and the "parabolic" behavior. However,
the $\beta$-FPU model does not obey the CV equation.

This paper continues the line of research started in \cite{p38} and investigates the relationship
between structure, parameters and macroscopic description of the nonstationary heat transfer in
models with on-site potential. Such models are known to have the normal heat conduction
\cite{p9,p10,p14}. We are going to check (a) whether the effects of hyperbolicity can be observed
in such models; (b) whether the thermal transport in these models can be described by linear
equations and (c) whether the lowest -- order CV equation is suitable for description of these phenomena.

\section{Models and procedures}
The stationary heat transfer is characterized by single coefficient of the heat (or thermal)
conductivity. In numeric experiments it is measured either by direct simulation of the stationary
flux under constant temperature gradient, or from autocorrelation of the heat flux in isolated
system via Green-Kubo formula \cite{p1}. The non-stationary conduction involves at least two
parameters (CV equation) or even more in more advanced models. Moreover, one should not bind
himself by a priori selection of the empiric equation to fit the results. Then, it is desirable
to find the characteristics of the process which can be measured directly from the numeric data
and are not related to specific choice of the macroscopic empiric equation.

One such choice may be observation of the temperature waves (the second sound), implemented,
for instance, in papers \cite{p27,p35,p36,p37}. We adopt more general approach, based on
relaxation profiles of different spatial modes of the temperature field. This approach is related
to early numeric simulations on the non-stationary heat conduction in argon crystals \cite{p30}
and of thermal conductivity in superlattices \cite{p34}.

In order to illustrate the approach, let us refer to 1D version of the CV equation for temperature:
\begin{equation}
\label{f2}
\tau \frac{\partial ^{2} T}{\partial t^{2}} +\frac{\partial T}{\partial t}
=\alpha\frac{\partial^{2} T}{\partial x^{2}}
\end{equation}

where $\alpha$ is the temperature conduction coefficient.

To solve this equation, let us consider the problem of non-stationary heat conduction in a
one-dimensional specimen with periodic boundary conditions $T(L,t)=T(0,t)$, where $T(x,t)$
is the temperature distribution, $L$ is the length of the specimen, $t>0$. If it is the case,
one can expand the temperature distribution into Fourier series:
\begin{equation}
\label{f3}
T(x,t)=\sum _{k=-\infty }^{\infty }a_{k} (t)\exp(2\pi ikx/L)
\end{equation}
with $a_{k} (t)=\bar{a}_{-k}(t)$, since $T(x,t)$ should be real, $k$
is the modal number.

Substituting (\ref{f3}) to (\ref{f2}), one obtains the equations for time evolution of the modal amplitudes:
\begin{equation}
\label{f4} \
\tau\ddot{a}_{k}+\dot{a}_{k}+\frac{4\pi ^{2}k^{2}\alpha }{L^{2}}a_{k} =0
\end{equation}

Solutions of Eq. (\ref{f4}) are written as:
\begin{equation}
\label{f5}
a_{k}(t)=C_{1k}\exp (\lambda_1t)+C_{2k}\exp(\lambda_2t),
\end{equation}
where $\lambda_{1,2}=(-1\pm \sqrt{1-16\pi^2 k^2\alpha\tau/L^2})/2\tau$,
$C_{1k}$ and $C_{2k}$ are constants determined by the initial distribution.

From (\ref{f5}) it immediately follows that for sufficiently short modes the temperature profile
will relax in oscillatory manner:
\begin{equation}
\label{f6}
a_{k}(t)\sim\exp(-t/2\tau)\exp(i\omega_{k}t),
\end{equation}
where $\omega_{k}=\sqrt{16\pi^2k^2\alpha\tau/L^2-1}/2\tau$,
$k>L/4\pi\sqrt{\alpha\tau}$.

If the specimen is rather long ($L\gg 4\pi\sqrt{\alpha\tau}$) then for small wavenumbers
(acoustical modes):
\begin{equation}
\label{f7}
\lambda_1\approx -1/\tau,~~\lambda_2\approx -4\pi^{2}k^{2}\alpha/L^{2}.
\end{equation}
The first eigenvalue describes fast initial transient relaxation, and the second one corresponds
to stationary slow diffusion and, quite naturally, does not depend on $\tau$.
This limit corresponds to Fourier heat conduction.

So, we can conclude that there exists a critical length of the mode
\begin{equation}
\label{f8}
l_0=4\pi \sqrt{\alpha\tau}.
\end{equation}
which separates between two different types of the relaxation: oscillatory and diffusive.

The oscillatory behavior is naturally related to the hyperbolicity of the system and is
qualitatively inconsistent with the Fourier law. Here the critical scale is derived from the CV
equation; however, it is clear that this scale, if it exists, can be revealed from numeric data
without relying on any particular macroscopic description. Moreover, this critical scale and its
dependence on the temperature and other parameters of the model help to understand how significant
are the deviations from parabolicity in given model. In particular, if the size of the system is
less than this critical mode length, the Fourier law cannot be used at all.

Careful analysis of the relaxation profiles for different modes of the
temperature perturbations also can give rather important insights.
For instance, Equation (\ref{f6}) predicts that all
oscillatory modes will decay with the same decrement related to unique relaxation time. One can
verify this prediction by comparison with the numeric data and thus to evaluate the accuracy
of the CV equation for particular model.

In order to measure the critical size $l_0$, the numeric experiment should be designed in order
to simulate the relaxation of thermal profile to its equilibrium value for different spatial modes.
For general one-dimensional model with on-site potential we simulate the chain of particles
with unit masses with Hamiltonian
\begin{equation}
\label{f9}
H=\sum _{n=1}^N[p_{n}^{2}/2+V(u_{n+1}-u_{n})+U(u_{n})].
\end{equation}
Here $u_n$ is the displacement of the $n$-th particle, $p_{n} =\dot{u}_{n}$, $V(\rho)$
is the potential of the nearest-neighbor interaction and $U(u)$ is the on-site potential
(the minimum of $V(\rho)$ corresponds to $\rho=0$ and minimum of $U(u)$ is at $u=0$).
The global minimum of the potential energy corresponds to an unperturbed state $\{u_{n}=0\}_{n=1}^N$.
Boundary conditions are adopted to be periodic.

In order to obtain the initial nonequilibrium
temperature distribution, all particles in the chain were initially connected to common Langevin thermostats.
For this sake, the following system of equations was simulated:
\begin{eqnarray}
\nonumber
\ddot{u}_{n}&=&V'(u_{n+1}-u_{n})-V'(u_{n}-u_{n-1})-U'(u_{n})\\
&&-\gamma\dot{u}_{n}+\xi_{n},~~n=1\dots N,
\label{f10}
\end{eqnarray}
where $\gamma$ is the damping coefficient and the white noise $\xi_n$ is normalized by the
following conditions:
\begin{equation}
\label{f11}
\langle\xi_{n}(t)\rangle=0,~~
\langle\xi_{n}(t_{1})\xi_{m}(t_{2})\rangle=2\gamma T_{n}\delta_{nm}\delta(t_{1}-t_{2}),
\end{equation}
where $T_n$ is the prescribed temperature of the $n$-th particle.

In order to study the relaxation of different spatial modes of the initial temperature
distribution, its profile has been prescribed as
\begin{equation}
\label{f12}
T_{n}=T_{0}+A\cos[2\pi(n-1)/Z],
\end{equation}
where $T_0$ is the average temperature, $A$ is the amplitude of the perturbation, and $Z$ is the
length of the mode (number of particles).
The overall length of the chain $N$ has to be multiple of $Z$ in order to ensure the periodic
boundary conditions.
After the initial heating in accordance with (\ref{f12}), the Langevin thermostat was removed and
relaxation of the isolated system to a stationary temperature profile is studied.
The results were averaged over about $10^6$ realizations of the initial profile $\{T_n\}_{n=1}^N$
in order to reduce the effect of fluctuations.

In this paper, we analyze three models with linear nearest-neighbor interaction, which differ by the
shape of the on-site potential. The Hamiltonian of all three models is written as:
\begin{equation}
\label{f13}
H =\sum_{n=1}^N[ p_{n}^{2}/2 +(u_{n+1}-u_{n})^{2}/2+U(u_{n})].
\end{equation}

Unit coefficient of the parabolic potential of the nearest-neighbor interaction does not
effect the generality. Real difference between the models appears due to different
choices of the on-site potential $U(u)$. We treat three topologically different on-site potentials:
sinusoidal potential (Frenkel-Kontorova  model, FK)
\begin{equation}
\label{f14}
U(u)=1-\cos u;
\end{equation}
$\phi^4$-  potential
\begin{equation}
\label{f15}
U(u)=2u^2(u-2\pi)^2/\pi^4;
\end{equation}
$\phi^4$+ potential
\begin{equation}
\label{f16}
U(u)=u^2/2+u^{4}/4.
\end{equation}

Substrate potentials (\ref{f14}), (\ref{f15}) and (\ref{f16}) differ topologically.
FK potential is periodic and bounded. Potential $\phi^4$- is double-well. To simplify the comparison,
 a distance between the well minima and a height of  the potential barrier are the same as for
the FK model. Potential $\phi^4$+ is single-well and unbounded.
Our goal is to check how these differences affect the nonequilibrium heat transfer. From the
viewpoint of equilibrium heat transfer, all these models have convergent thermal conduction
coefficient in the thermodynamical limit \cite{p1}.
%---------------------------- Fig. 1 ------------------------------------
\begin{figure}[tbp]
\includegraphics [angle=0, width=1\linewidth]{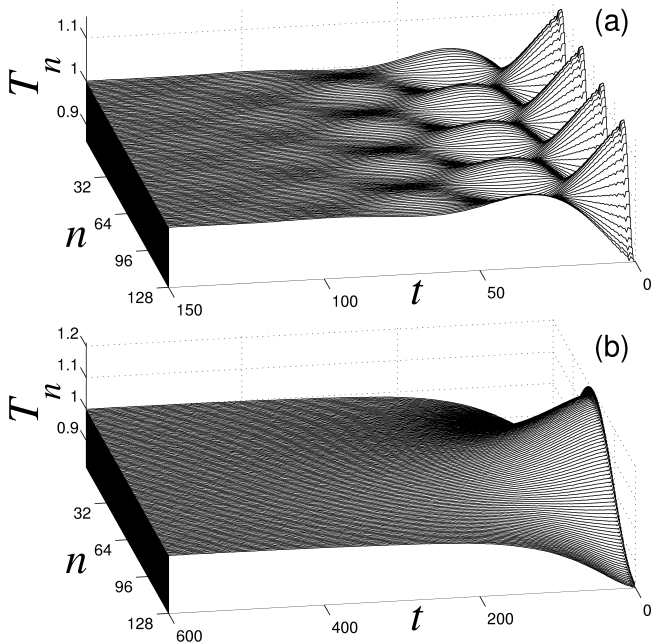}
\caption{\label{fig1}%\protect
Relaxation of initial periodic thermal profile in the chain with FK potential,
$N=1024$, $T_0=1$, $A=0.2$ (a) $Z=32$ and (b) $Z=128$. Only part of the chain
for $n$ from 1 to 128 is demonstrated.
}
\end{figure}
%---------------------------- Fig. 1 ------------------------------------

\section{Results}
The first set of simulations reported here has been devoted to verification of transition from
oscillatory to smooth relaxation profile of the temperature perturbation for different initial wavelength.
Typical result of the simulation is presented in Fig. \ref{fig1}.
The FK chains with the same length $N=1024$, average temperature $T_0=1$ and amplitude of the
perturbation $A=0.2$
demonstrate qualitatively different relaxation profiles for different modal wavelength $Z$.
%---------------------------- Fig. 2 ------------------------------------
\begin{figure}[tbp]
\includegraphics[angle=0, width=1\linewidth]{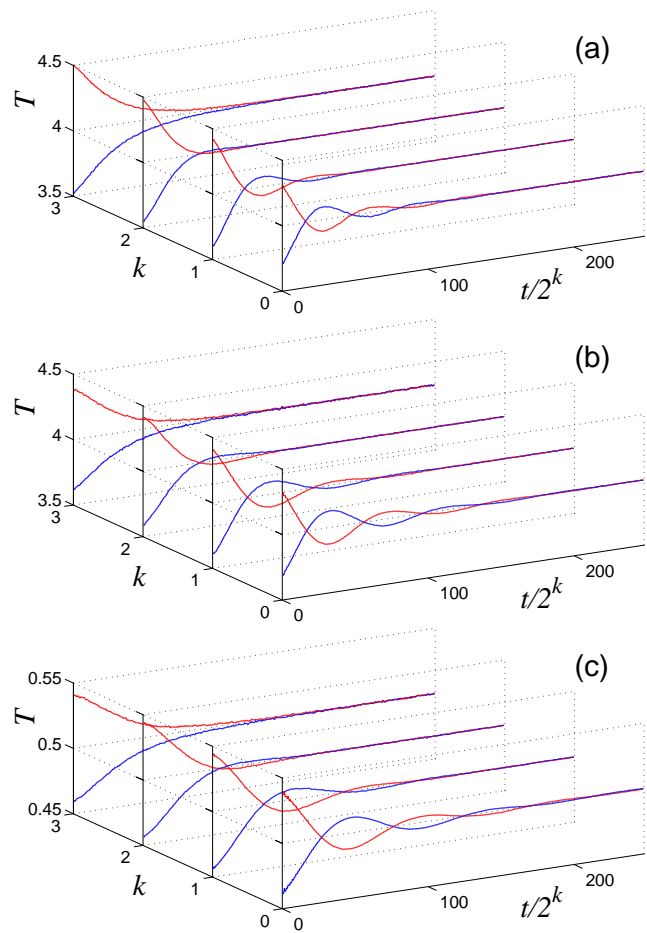}
\caption{\label{fig2}\protect
(Color online) Evolution of the relaxation profile in the chain (length $N=800$)
for $Z=50$, 100, 200 and 400 ($k=0$, 1, 2 and 3) with (a) FK potential ($T_0=4$,
$A=0.5$); (b) $\phi^4$- potential ($T_0=4$, $A=0.5$); (c) $\phi^4$+ potential
($T=0.5$, $A=0.05$).
Time dependence of the mode maximum $T(1+Z/2)$ [red (gray) lines] and minimum
$T(1)$ [blue (black) lines] are depicted. Note that the time scales are different
for each curve in order to fit them at one figure.
}
\end{figure}
%---------------------------- Fig. 2 ------------------------------------

Results of the simulations presented in Fig. \ref{fig2} demonstrate that all three investigated
models with the on-site potential clearly exhibit similar transition from diffusive to oscillatory decay
pattern as the characteristic wavelength decreases.
It is possible to distinguish qualitatively three different relaxation patterns: diffusive
(curves with $k=3$), oscillatory (curves with $k=0$, 1) and  crossover (curves with $k=2$).
In other terms, for all these models there exists the critical
wavelength scale $l^*$ which separates the oscillatory from the diffusive decay.
For the simulations presented here this characteristic scale is
\begin{equation}
\label{f17} l^* \sim 200\div 300.
\end{equation}

Thus, already at this step one can conclude that the "hyperbolic" behavior can be observed
in all three models.

\begin{figure}[tbp]
\includegraphics[angle=0, width=1\linewidth]{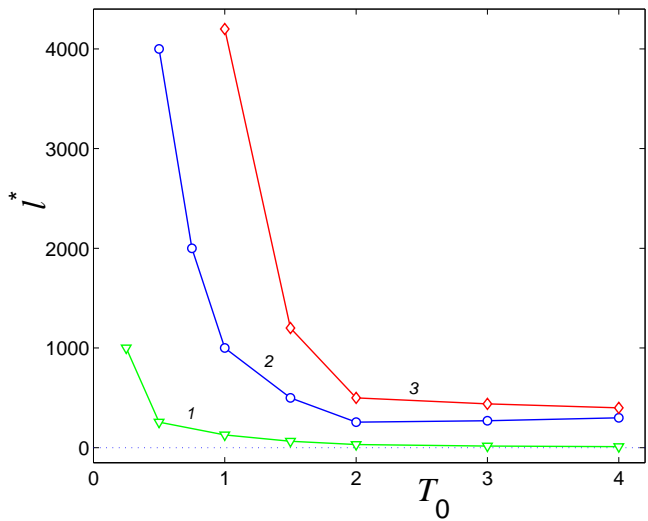}
\caption{\label{fig3}%\protect
Dependence of the critical wavelength $l^*$ on average temperature $T_0$
of the chain with $\phi^4$+ substrate (curve 1), FK substrate (curve 2)
and $\phi^4$- substrate potential (curve 3).
}
\end{figure}

The wide range obtained in (\ref{f17}) is due to relatively small chain length; this simulation is
sufficient to reveal the existence of transition between the decay patterns, but not sufficient
for more or less exact estimation of the critical wavelength scale. Still, one can mention that
similar values of the critical scale in Fig. \ref{fig2} were obtained for different values of the
average temperature of the system. Thus, one can conjecture that $l^*$ should be dependent on the
temperature -- similar observation was made also for the models with conserved momentum \cite{p38}.
To verify this conjecture, we have performed more exact measurements of $l^*$ (for longer chains,
up to 10000 particles, and with higher resolution) for all three models.
The results are presented in Fig. \ref{fig3}. As the temperature decreases,
the crossover length $l^*$ increases monotonously.
%---------------------------- Fig. 3 ------------------------------------
%---------------------------- Fig. 3 ------------------------------------

\section{Linearity of the thermal relaxation profiles}

\begin{figure}[tbp]
\includegraphics[angle=0, width=1\linewidth]{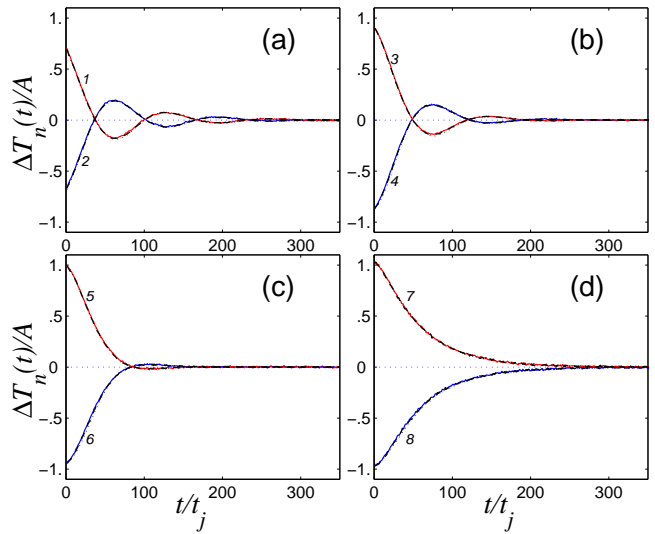}
\caption{\label{fig4}%\protect
(Color online)
Dependence of rescaled temperature amplitude $\Delta T_n(t)/A=(T_n(t)-T_0)/A$
on rescaled time $t/t_j$ for FK chain: (a) $Z=50$, $t_j=0.5$; (b) $Z=100$,
$t_j=1$;  (c) $Z=200$, $t_j=2$;  (d) $Z=400$, $t_j=4$.
The length $N=800$, average temperature $T_0=4$.
Lines 1,3,5,7 [red (gray)] correspond to the maximum of the temperature in the chain $n=1$,
and Lines 2,4,6,7 [blue (black)] -- to the minimum $n=Z/2+1$.
Solid lines correspond to the perturbation amplitude $A=0.5$, dashed lines -- to $A=1$, dots --
to $A=2$.  The time scales are different for each curve in order to fit them at one figure.
}
\end{figure}

In order to check to what extent it might be possible to assume that the macroscopic equations
of the nonstationary heat conduction for given models are linear, we have
simulated the evolution of initial temperature distribution (\ref{f12}) with varying amplitude of
the perturbation $A$, keeping all other parameters fixed. Then, the value
$\Delta T_n(t)=(T_n(t) -T_0)/A$ is plotted versus time. If the results coincide
for different values of $A$, then it is possible to conclude that the relaxation
of the thermal perturbations can be described by linear equation.
In Fig. \ref{fig4} the simulation results for the FK model are demonstrated.
%---------------------------- Fig. 4 ------------------------------------
%---------------------------- Fig. 4 ------------------------------------

For all regimes of the relaxation, one barely sees any difference for the relaxation profiles
with different initial amplitudes; it happens despite the fact that the perturbation is by no
means small -- it achieves the half of the average temperature. Quite similar results were
obtained also for $\phi^4$- chain and $\phi^4$+ chains and are omitted for the sake of brevity.
So, one can conclude that the process of thermal relaxation in all these models may be described
macroscopically with the help of linear equations with rather good accuracy. Similar conclusion
was achieved also for the systems with conserved momentum, but with somewhat
different method \cite{p38}.

This conclusion does not seem to be trivial. The models under consideration are essentially
nonlinear and it is not completely clear why linear macroscopic equations should be so
suitable for the description of the thermal relaxation, especially for relatively large perturbation amplitudes.

\section{Thermal relaxation in oscillatory regime and the CV approximation}
As it was mentioned before, the Cattaneo-Vernotte law predicts that the thermal relaxation will be
exponential and that the relaxation time will be the same for all oscillatory modes.
Both these statements are not self - evident and should be verified numerically.

For the sake of comparison, let us consider the same process of the relaxation of thermal waves
in an infinite linear chain with the on-site potential. Such chain will be described by Equation (\ref{f13})
with the external potential
\begin{equation}
\label{f18}
U_{L} (u)=\omega^{2}u^{2}/2.
\end{equation}

Appropriate equations of motion will take the form
\begin{equation}
\label{f19}
\ddot{u}_{n}+2u_{n}-u_{n-1}-u_{n+1}+\omega^{2}u_{n}=0.
\end{equation}
For the sake of simplicity, let us adopt that the initial temperature distribution is realized
through attribution of initial velocities to each particle, with zero initial displacements.
By virial theorem, such choice will not affect the long-time asymptotics. Namely, the initial
conditions for Equation (\ref{f19}) are formulated as:
\begin{eqnarray}
\label{f20}
u_{n}(0)=0,~~\dot{u}_{n}(0)=\eta_{n},~~\langle\eta_{n}\rangle =0,\\
\langle\eta_{n}\eta_{m}\rangle =2[T_{0}+A\cos(2\pi n/Z)]\delta_{nm}.
\nonumber
\end{eqnarray}
where $\eta_n$ is random value with Gaussian distribution according to (\ref{f20}).
The averaging is performed over the ensemble of the initial conditions. By standard transformations,
it is easy to obtain the following solution of (\ref{f19}):
\begin{eqnarray}
\label{f21}
\dot{u}_{n}(t)=\sum_{m=-\infty }^{\infty }\eta_{m}G(n-m,t), \\
G(p,t)=\frac{1}{\pi}\int_{0}^{\pi }\cos(t\sqrt{\omega^{2}+4\sin^{2}\mu}-2p\mu)d\mu.
\nonumber
\end{eqnarray}

The temperature distribution at arbitrary moment of time may be presented as:
\begin{eqnarray}
\label{f22}
T_{n}(t)=\langle\dot{u}_{n}^{2}\rangle=T_{0}+2A g_{Z}(t)\cos(2\pi n/Z), \\
g_{Z}(t)=\sum_{p=-\infty }^{\infty} G^{2}(p,t)\cos(2\pi p/Z).
\nonumber
\end{eqnarray}

Time dependence of given thermal mode is completely governed by long-time asymptotics of function
$g_{Z}(t)$. For given linear model with on-site potential it is difficult to perform the exact
integrations and summations in equations (\ref{f21}), (\ref{f22}) (for the case without the
on-site potential these expressions can be reduced to compact exact expressions in terms
of Bessel functions). Instead, we are going to
derive the required asymptotics by approximate method, taking into account only generic
features of the dispersion relation used in (\ref{f21}). For this sake, let us consider the
general Green function similar to (\ref{f21}):
\begin{equation}
\label{f23}
G_{0}(p,t)=\frac{1}{\pi}\int_{0}^{\pi }\cos(tf(\mu)-2p\mu)d\mu,
\end{equation}
where $f(\mu)$ is bounded at interval $(0,\pi)$ and has a single inflection point $\mu^*$ with
positive first derivative, which satisfies the conditions
\begin{equation}
\label{f24}
f''(\mu^{*})=0,~~f'(\mu^{*})>0,~~\mu^{*}\in [0,\pi ].
\end{equation}
Then, for $t$ large enough, the integral in (\ref{f23}) can be evaluated by a stationary phase
method. The condition of the stationary phase for specific values of $p$, $t$ is presented as:
\begin{equation}
\label{f25}
\Omega(\mu,p,t)=tf(\mu)-2p\mu,~~\Omega'(\mu_{0})=tf'(\mu_{0})-2p=0.
\end{equation}

For the sake of simplicity, we will consider only the case where (\ref{f25}) has unique solution.
In the vicinity of this point, the phase is expanded as:
\begin{eqnarray}
\label{f26}
\Omega =f(\mu_{0})t-2p\mu_{0}+\frac{1}{2}f''(\mu_{0})t(\mu -\mu_{0})^2\\
+\frac{1}{6}f'''(\mu_{0})t(\mu -\mu_{0})^{3}+\dots.
\nonumber
\end{eqnarray}
Standard evaluation of integral (\ref{f23}) for the case $t\gg p$ will, therefore, yield
\begin{equation}
\label{f27}
G_{0}(p,t)\sim t^{-1/2},~~t\to\infty.
\end{equation}

Estimation (\ref{f27}) is, however, wrong in the vicinity of the inflection point, which is
defined by condition $\mu_{0}=\mu^{*}$. In this case, the term with the second derivative
in expansion (\ref{f26}) will disappear and one will obtain:
\begin{equation}
\label{f28}
\Omega =f(\mu^{*})t-2p\mu^{*}+\frac{1}{6}f'''(\mu^{*})t(\mu -\mu^{*})^{3}+\dots.
\end{equation}
Quite obviously, evaluation of integral (\ref{f23}) with phase expansion (\ref{f28})
will yield:
\begin{eqnarray}
\nonumber
G_{0}(p,t)=\frac{1}{\pi}\int_{0}^{\pi }\cos[\Omega(\mu ,p,t)]d\mu\\
\sim t^{-1/3}\cos[f(\mu^{*})t-2p\mu^{*}+\psi ]+O(t^{-2/3}),
\label{f29}
\end{eqnarray}
where $\psi$ denotes constant phase shift. Expansion (\ref{f29}) is valid for
$2p/t\approx f'(\mu^{*})$, which, in turn, corresponds to maximum group velocity
of linear waves in the
system under consideration. Estimation (\ref{f29}) suggests, more exactly,
that the expansion is valid for the interval of p defined as:
\begin{equation}
\label{f30}
|p-p^{*}|\sim (p^{*})^{1/3},~~p^{*} =tf'(\mu^{*})/2.
\end{equation}

Due to slower decrease rate with respect to time, the terms in expansion (\ref{f22}) in the
interval of index defined by estimation (\ref{f30}) will be the most significant for long-time
behavior. Then, the sum in (\ref{f22}) may be estimated as:
\begin{equation}
\label{f31}
g_{k}(t)\sim t^{-2/3}\sum_{p=p^{*}-(p^{*})^{1/3}}^{p^{*}+(p^{*})^{1/3}}
\cos(\frac{2\pi p}{Z})\cos^{2}[f(\mu^{*})t-2p\mu^{*}+\psi].
\end{equation}

For the most interesting case of relatively small wavenumbers the input of
the sum in (\ref{f31}) will be of order of the square root of the number of participating
summands, i.e. of order $(p^{*})^{1/6}$. Consequently, with account of (\ref{f30}), one finally
arrives to the following estimation:
\begin{equation}
\label{f32}
g_{Z}(t)\sim t^{-2/3}(p^{*})^{1/6} \sim t^{-1/2}.
\end{equation}

\begin{figure}[tbp]
\includegraphics[angle=0, width=1\linewidth]{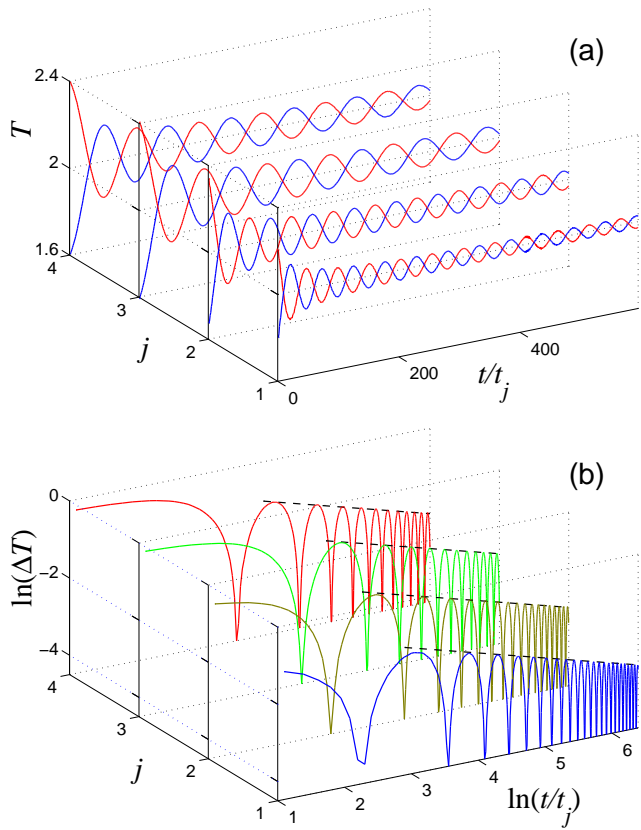}
\caption{\label{fig5}%\protect
(Color online) (a) Evolution of the relaxation profile in the linear chain
(frequency $\omega=1$, length $N=1024$, average temperature $T_0=2$, amplitude $A=0.4$)
for $Z=32$, 64, 128 and 256 ($j=1$, 2, 3 and 4; time scales $t_1=1$, $t_2=1.5$, $t_3=2$
and $t_4=4$ respectively).
Time dependence of the mode maximum $T_{1}$ [red (gray) lines] and minimum
$T_{Z/2+1}$  [blue (black) lines] are depicted.
The time scales are different for each curve in order to fit them at one figure.
(b) The same results are presented in a double logarithmic coordinates,
$\Delta T(t)=|T_1(t)-T_0|+|T_{Z/2+1}(t)-T_0|$. Dashed fitting lines correspond
to the power law $\Delta T \sim t^{-0.48}$.
}
\end{figure}

As it is clear from the treatment, estimation (\ref{f32}) should be rather generic for linear chains.
This power-law decay of the thermal mode is illustrated in Fig. \ref{fig5},
%---------------------------- Fig. 5 ------------------------------------
%---------------------------- Fig. 5 ------------------------------------

Therefore, we see that the thermal "relaxation" in linear chains obeys power law rather than exponential.
One can also mention that, quite as expected, the long-time asymptotics is not effected by the
difference between the finite chain with periodic boundary conditions used for the simulations
and the infinite chain used treated analytically.
Also, it is not very surprising that the linear chains do not obey the CV law.  However, one can
expect that the nonlinear chains with on-site potential, known to have normal heat conduction
\cite{p1} will obey it to some extent. At least, the results for the models with conserved
momentum \cite{p38} suggest such conjecture.

\begin{figure}[tbp]
\includegraphics[angle=0, width=1\linewidth]{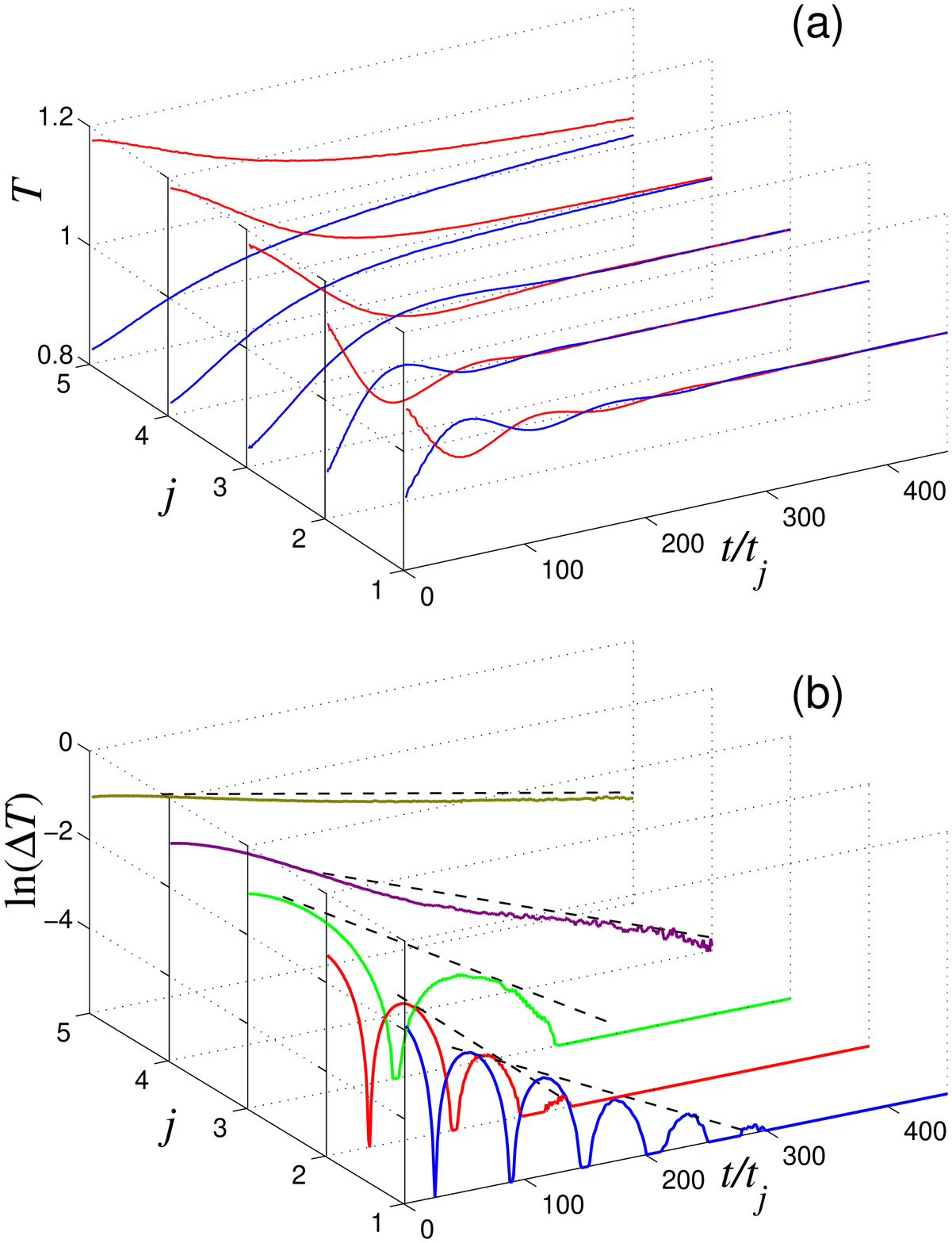}
\caption{\label{fig6}%\protect
(Color online) (a) Evolution of the relaxation profile in the $\phi^4+$ chain
(length $N=1024$, average temperature $T_0=1$, amplitude $A=0.2$)
for $Z=16$, 32, 64, 128 and 256 ($j=1$, 2, 3, 4 and 5; time scale $t_1=0.25$, $t_2=t_3=0.5$, $t_4=1$,
and $t_5=2$ respectively).
Time dependence of the  maximum $T_{1}$ [red (gray) lines] and minimum
$T_{Z/2+1}$ [blue (black) lines] are depicted. The time scales are different for
each curve in order to fit them at one figure.
(b) Decay profile of the temperature difference
$\Delta T(t)=|T_1(t)-T_0|+|T_{Z/2+1}(t)-T_0|$ in semilogarithmic coordinates.
Dashed lines correspond to $\Delta T\sim \exp(-t/t_r)$
with $t_r=18.9$, 22.2, 31.3, 100 and 370 for $j=1$, 2, 3, 4 and 5 respectively.
}
\end{figure}

The first statement to check is whether the decay of the thermal disturbances in the considered
nonlinear models is indeed exponential. For this sake, we simulate the relaxation for different
wavenumbers $k$  for $\phi^4+$ on-site potential (\ref{f16}). The results are presented in Fig. \ref{fig6}.
%---------------------------- Fig. 6 ------------------------------------
%---------------------------- Fig. 6 ------------------------------------
This simulation clearly demonstrates the exponential character of decay for
$\Delta T(t)=|T_1(t)-T_0|+|T_{Z/2+1}(t)-T_0|$:
$$
\Delta T(t)\sim \exp(-t/t_r),~~~\mbox{when}~~t\rightarrow\infty,
$$

\begin{figure}[tbp]
\includegraphics[angle=0, width=1\linewidth]{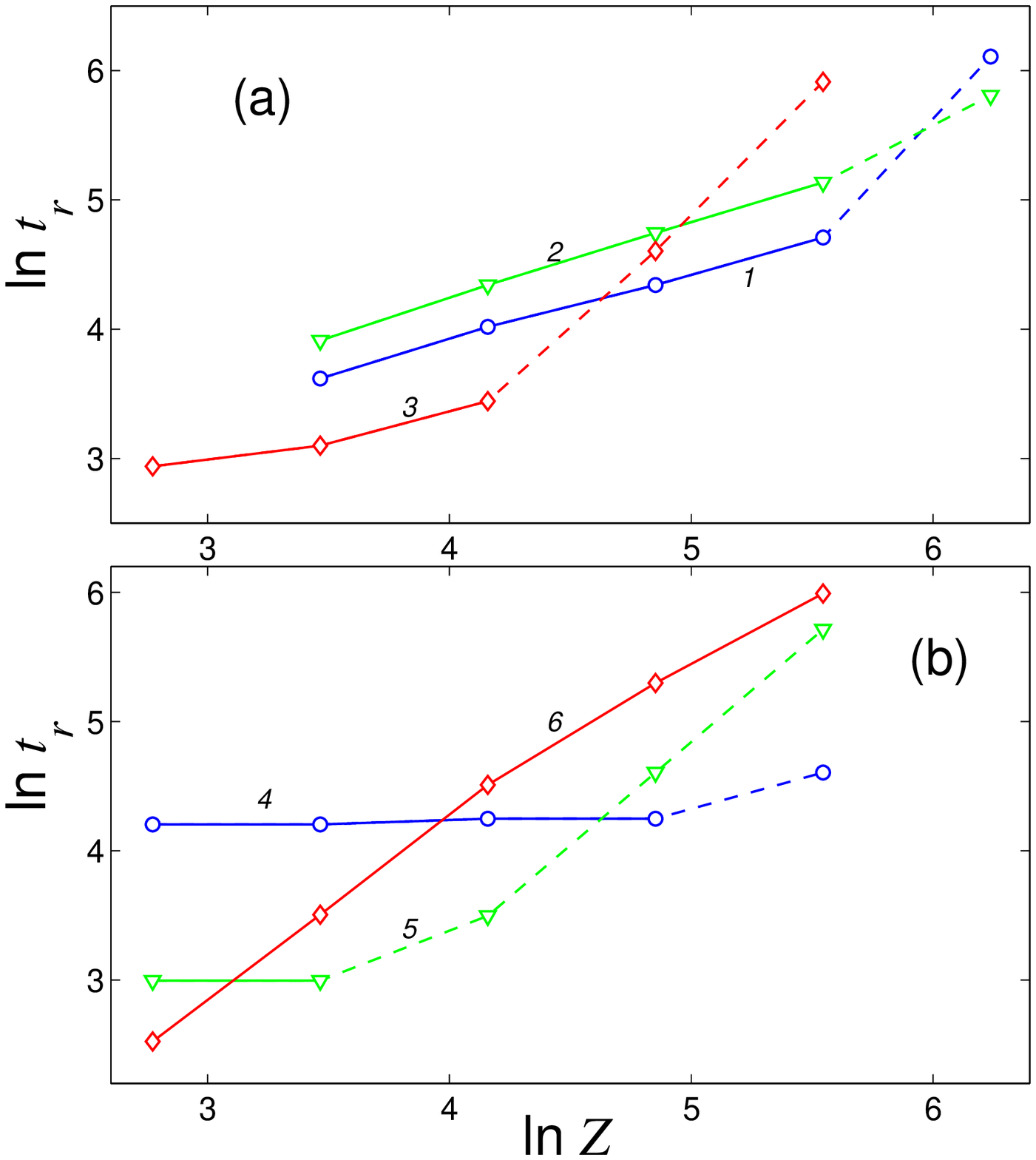}
\caption{\label{fig7}%\protect
(Color online)
Dependence of the characteristic relaxation time $t_r$ on the period of initial thermal
perturbation $Z$ (double logarithmic scale):
 (a) the models with on-site potential: FK model (line 1,
average temperature $T_0=2$, perturbation amplitude $A=0.5$),
$\phi^4-$ model (line 2, $T_0=2$, $A=0.5$), $\phi^4+$ model (line 3, $T_0=1$, $A=0.2$);
(b) the models with conserved momentum \cite{p38}: chain of rotators (line 4, $T_0=0.5$, line 5, $T_0=0.3$)
FPU (line 6, $T_0=20$). For all models, the solid part of the line corresponds to the oscillatory
relaxation pattern and the dashed -- to the crossover and monotonous relaxation.
}
\end{figure}

However in Fig. \ref{fig6} we already can see that the characteristic relaxation time $t_r$
in the oscillatory regime depends on the perturbation wavelength, clearly at odds with the CV law (comp.(\ref{f6})). To further clarify this point,
we present in Fig. \ref{fig7} the dependence of the characteristic relaxation time on the thermal
perturbation wavelength for all three models with the on-site potentials studied above,
as well as for the models with conserved momentum (FPU and chain of rotators) studied in \cite{p38}.
%---------------------------- Fig. 7 ------------------------------------
%---------------------------- Fig. 7 ------------------------------------

As we can see from these results, three out of five models (FPU, FK and $\varphi^{4}$-) exhibit peculiar
scaling of the relaxation time in the oscillatory regime, which conforms to the approximate law
\begin{equation}
\label{f32}
\tau_{k} \sim (N/k)^{\beta},
\end{equation}
with $\beta \approx 0.45$ for the FK and $\varphi^{4}$- and $\beta \approx 1.4$  for the FPU model.
As for $\varphi^{4}$+ model, the relaxation time also depends on the wavelength
of the initial thermal profile, but it is difficult to speak about some peculiar scaling.
Only in the chain of rotators the relaxation time in the oscillatory regime almost does not
depend on the wavelength of the temperature perturbation.

\section{Concluding remarks}
On the basis of the simulations presented above, one can conclude that the one-dimensional models
with topologically different on-site potentials exhibit qualitatively similar behavior with
respect to nonstationary thermal conductivity. The thermal relaxation can be quite accurately described by linear equations even for relatively high perturbation amplitudes. All three models
demonstrate transition from oscillatory to diffusive relaxation regimes for growing wavelength
of the initial thermal profile. Characteristic crossover wavelength rapidly decreases with the
temperature increase.

Qualitatively, such behavior is compatible with the Cattaneo-Vernotte equation. However,
quantitative study reveals that no unique relaxation time exists for different spatial harmonics
of the initial temperature profile. For part (but not all) of the models, the relaxation times
in the oscillatory regime approximately obey scaling law with respect to the wavelength of the initial
thermal profile.

All studied models are known to obey the Fourier law in the thermodynamical limit. The fact that
the CV law is not suitable for description of the nonstationary heat transfer in these systems is
rather surprising. Moreover, the scaling in FK and $\varphi^{4}$- resembles
that in the FPU model, which has divergent heat conduction, although the scaling exponents are different. Still, at this stage one cannot claim that these exponents are in any sense universal - a lot of further analysis is required. The chain of rotators turns out to be  "exceptional" -- it seems to obey the CV law both for the long and the short wavelengths of the temperature distribution.

Scaling relationship (\ref{f32}) suggests that the macroscopic equations describing the nonstationary
heat conduction in these models should include fractional derivatives. It is quite unusual that such "fractional' terms will appear as "hyperbolic" corrections to common Fourier law.


\begin{references}

\bibitem{p1}
S. Lepri, R. Livi and A. Politi,
%Thermal Conduction in Classical Low-Dimensional Lattices,
Phys. Rep. {\bf 377}, 1 (2003).

\bibitem{p2}
F. Bonetto, J. L. Lebowitz and L. Rey-Bellet,
Fourier Law: A Challenge to Theorists, Mathematical Physics 2000,
A. S. Fokas, Imperial College Press, p. 128 (2000).

\bibitem{p3}
R. E. Peierls,
Quantum Theory of Solids,
Oxford University Press, London, 1955.

\bibitem{p4}
E. Fermi, J. Pasta and S. Ulam,
Studies of Non Linear Problems, American Mathematical Monthly,
{\bf 74}, Issue 1, 1967 (Reprint from Los Alamos Reports, 1955).

\bibitem{p5}
S. Lepri, R. Livi and A. Politi,
%On The Anomalous Thermal Conductivity of One-Dimensional Lattices,
Europhysics Letters {\bf 43}, 271 (1998).

\bibitem{p6}
B. Hu, B. Li and  H. Zhao,
%Heat Conduction in One-Dimensional Non-Integrable Systems,
Phys. Rev. E {\bf 61}, 3828 (2000).

\bibitem{p7}
C. Giardina, R. Livi, A. Politi and M. Vassalli,
%Finite Thermal Conductivity in 1D Lattices,
Phys. Rev. Lett. {\bf 84}, 2144 (2000).

\bibitem{p8} O. V. Gendelman and A. V. Savin,
%Normal Heat Conductivity of the One-Dimensional Lattice with Periodic Potential
%of Nearest-Neighbor Interaction,
Phys. Rev. Lett. {\bf 84}, 2381 (2000).
%
%O.V.Gendelman and A.V.Savin,
%Heat Conductivity in the Chain of Rotators,
%\textit{Solid State Physics}, v.43,  355-364, 2001

\bibitem{p9}
A. V.  Savin and O. V. Gendelman,
%Heat Conduction in One-Dimensional Lattices with on-site Potential,
Phys. Rev. E {\bf 67}, 041205 (2003).

\bibitem{p10}
O. V.  Gendelman and A. V. Savin,
%Heat Conduction in One-Dimensional chain of Hard Discs with Substrate Potential,
Phys. Rev. Lett. {\bf 92},  074301 (2004).

\bibitem{p11}
E. Pereira and R. Falcao,
%Heat Conduction in a Chain with a Weak Interparticle Anharmonic Potential,
Phys. Rev. Lett. {\bf 96}, 100601 (2006).

\bibitem{p12}
G. Santhosh and D. Kumar,
%Universality of thermal conduction in vibrating chains for a class of potentials,
Phys. Rev. E {\bf 77}, 011113 (2008).

\bibitem{p13}
A. V. Savin, G. P. Tsironis and X. Zotos,
%Thermal Conductivity of a Classical One-Dimensional Spin-Phonon System,
Phys. Rev. B {\bf 75}, 214305 (2007).

\bibitem{p14}
B. Hu and L. Yang,
%Heat conduction in Frenkel-Kontorova Model,
Chaos {\bf 15}, 015119 (2005).

\bibitem{p15}
K. Saito and A. Dhar,
%Heat Conduction in a Three Dimensional Anharmonic Crystal,
Phys. Rev. Lett. {\bf 104}, 040601 (2010).

\bibitem{p16}
H. Shiba and N. Ito,
%Anomalous heat conduction in three-dimensional nonlinear lattices,
Journal of the Physical Society of Japan {\bf 77}, 054006 (2008).

\bibitem{p17}
D. D. Joseph and L. Preziosi,
%Heat Waves,
Reviews of Modern Physics {\bf 61}, 41 (1989).

\bibitem{p18}
D. D. Joseph and L. Preziosi,
Addendum to "Heat Waves",
Reviews of Modern Physics {\bf 62}, 375 (1990).

\bibitem{p19}
D. S. Chandrasekhararaiah,
Appl. Mech. Rev. {\bf 39}, 355 (1986).

\bibitem{p20}
D. S. Chandrasekhararaiah
Appl. Mech. Rev. {\bf 51}, 705 (1998).

\bibitem{p21}
P. Heino,
Journal of Comput. and Theor. Nanoscience {\bf 4}, 896 (2007).

\bibitem{p22}
D. Jou, J. Casas-Vazguez and G. Lebon,
Extended Irreversible Thermodynamics, 4th edition, Springer, 2010.

\bibitem{p23}
C. Cattaneo,
%A Form of Heat Conduction Equation which Eliminates the Paradox of Instantaneous Propagation,
Comp. Rend. {\bf 247}, 431, (1958).

\bibitem{p24}
P. Vernotte,
%Les Paradoxes de placeplacela Theorie Continue de L'equation de placela Chaleur,
Comp. Rend. {\bf 246}, 3154, (1958).

\bibitem{p25}
P. Antaki,
%Hotter than you think,
Machine Design {\bf 67}, 116 (1995).

\bibitem{p26}
C. I. Christov and P. M. Jordan,
%Heat Conduction Paradox Involving Second-Sound Propagation in Moving Media,
Phys. Rev. Lett. {\bf 94}, 154301 (2005).

\bibitem{p27}
J. Shiomi and S. Maruyama,
%Non Fourier Heat Conduction in a Single-Walled Carbon Nanotube: Classical Molecular Dynamics Simulations,
Phys. Rev. B {\bf 73},  205420 (2006).

\bibitem{p28}
S. T. Huxtable, D. G. Cahill, S. Shenogin, L. Xue, R. Ozisik, P. Barone, M. Usrey, M. S. Strano,
G. Siddons, M. Shim, and P. Keblinski,
%Interfacial heat flow in carbon nanotube suspensions,
Nat. Mater. {\bf 2}, 731 (2003).

\bibitem{p29}
D. H. Tsai and R. A. MacDonald,
%Molecular-dynamical study of second sound in a solid excited by a strong heat pulse,
Phys. Rev. B {\bf 14}, 4714 (1976).

\bibitem{p30}
S. Volz, J. B. Saulnier, M. Lallemand, B. Perrin, P. Depondt, and M. Mareschal,
%Transient Fourier-Law Deviation by Molecular Dynamics in Solid Argon,
Phys. Rev. B {\bf 54}, 340 (1996).

\bibitem{p31}
D. Tang and N. Araki,
%Microscopic Study on Non-Fourier Heat Conduction,
Thermophys. Prop. {\bf 22}, 16 (2001).

\bibitem{p32}
B. Y. Cao and Z. Y. Guo,
%Equation of motion of a phonon gas and non-Fourier heat conduction,
Journal of Applied Physics {\bf 102}, 053503 (2007).

\bibitem{p33}
S.G. Volz,
%Thermal Insulating Behavior in Crystals at High Frequencies,
Phys. Rev. Lett. {\bf 87}, 074301 (2001).

\bibitem{p34}
B. C. Daly, H.J. Maris, K. Imamura and S. Tamura,
%Molecular dynamics calculation of the thermal conductivity of superlattices,
Phys. Rev. B. {\bf 66}, 024301 (2002).

\bibitem{p35}
S. Flach and G. Mutschke,
Phys. Rev. E {\bf 49}, 5018 (1994).

\bibitem{p36}
T. Schneider and E. Stoll,
Phys. Rev. B {\bf 18}, 6468 (1978).

\bibitem{p37}
F. Piazza and S. Lepri,
%Heat Propagation in a Nonlinear Chain,
Phys. Rev. B {\bf 79}, 094306 (2009).

\bibitem{p38}
O. V. Gendelman and A. V. Savin,
%Nonstationary heat conduction in one-dimensional chains with conserved momentum,
Phys. Rev. E {\bf 81}, 020103(R) (2010).

\end{references}
\end{document}